\documentclass[conference]{IEEEtran}
\IEEEoverridecommandlockouts
\usepackage{cite}
\usepackage{amsmath,amssymb,amsfonts}
\usepackage{algorithmic}
\usepackage{graphicx}
\usepackage{textcomp}
\usepackage{xcolor}
\usepackage{url}
\usepackage{booktabs}

\def\BibTeX{{\rm B\kern-.05em{\sc i\kern-.025em b}\kern-.08em
    T\kern-.1667em\lower.7ex\hbox{E}\kern-.125emX}}

\begin{document}

\title{ASR for Affective Speech: Investigating Impact of Emotion and Speech Generative Strategy
}
\author{
\IEEEauthorblockN{Ya-Tse Wu}
\IEEEauthorblockA{
\textit{Department of Electrical Engineering,}\\
\textit{National Tsing Hua University}\\
Hsinchu, Taiwan \\
crowpeter@gapp.nthu.edu.tw}
\and
\IEEEauthorblockN{Chi-Chun Lee}
\IEEEauthorblockA{
\textit{National Tsing Hua University}\\
Hsinchu, Taiwan \\
cclee@ee.nthu.edu.tw}
}

\maketitle

\begin{abstract}
This work investigates how emotional speech and generative strategies affect ASR performance. We analyze speech synthesized from three emotional TTS models and find that substitution errors dominate, with emotional expressiveness varying across models. Based on these insights, we introduce two generative strategies: one using transcription correctness and another using emotional salience, to construct fine-tuning subsets. Results show consistent WER improvements on real emotional datasets without noticeable degradation on clean LibriSpeech utterances. The combined strategy achieves the strongest gains, particularly for expressive speech. These findings highlight the importance of targeted augmentation for building emotion-aware ASR systems.
\end{abstract}


\section{Introduction}
Automatic speech recognition (ASR) is a foundational technology in speech-based applications and plays a key role in modern human-computer interaction \cite{nguyen2025speech}. For real-world deployment, robustness is a critical requirement. While much of the existing research has focused on external factors such as background noise and reverberation \cite{6732927}, internal factors related to the speaker, particularly emotional expression, have received comparatively less attention despite their substantial impact on speech characteristics \cite{munot2019emotion}. Emotion plays a central role in human communication, not only affecting prosody, articulation, and voice quality but also characterized either by discrete categories (e.g., angry, happy, sad) or by continuous dimensions such as valence and arousal \cite{laukka2005dimensional}. Speech produced under different emotional states can vary considerably in both acoustic and linguistic patterns, posing challenges to conventional ASR systems \cite{batliner2010automatic}. Such variability poses challenges to conventional ASR systems, underscoring the need for models that can accurately recognize emotional speech and remain reliable under diverse real-world conditions.


Emotion has traditionally been regarded as a source of noise in the ASR process. For example, Catania et al. showed that emotional speech negatively impacts the performance of speech recognition and user input understanding, framing emotion as a disruptive factor in human-AI interaction \cite{catania2019automatic}. Other studies have also noted lexical and acoustic differences across emotional states. Gulmira et al. proposed a low-resource emotional speech recognition method that detects emotions based on a codebook of emotionally charged words in Kazakh language \cite{bekmanova2022emotional}. These findings suggest that emotional speech is more difficult to recognize than neutral speech due to its high variability in acoustic and prosodic features. Moreover, the high cost of collecting labeled emotional speech makes emotional TTS a scalable and low-cost alternative for training emotion-aware ASR systems.

One practical approach to addressing this challenge is through data augmentation, a technique widely used to improve robustness in noise-prone ASR systems \cite{hu2018generative, pervaiz2020incorporating}. Among these, controllable text-to-speech (TTS) models have emerged as a scalable tool for simulating diverse speaking conditions. Previous studies have shown that TTS-based augmentation can improve ASR performance for dysarthric speech \cite{leung24_interspeech}, suggesting that synthetic speech can help bridge the domain gap when real-world data is limited. In this context, emotional TTS offers a promising avenue for generating emotional speech that may help improve ASR performance under affective conditions.

Recent advances in large language model-based TTS systems, such as CosyVoice2 \cite{du2024cosyvoice} and EmoVoice \cite{yang2025emovoice}, have made it feasible to generate highly controllable and emotionally rich synthetic speech. However, the use of synthesized emotional speech for ASR training presents several challenges. Synthesized data may introduce acoustic or prosodic artifacts that distort recognition, and the emotional salience of synthetic speech may not fully align with real-world emotional expression. Additionally, it remains unclear which types of ASR errors are most impacted by emotional content, and how different TTS models influence these error patterns has not been systematically studied.

To better understand these effects, we conduct a series of analytical experiments to examine how synthetic emotional speech affects ASR performance. Starting from synthesized data, we analyze the predominant recognition errors introduced by an ASR model. Based on these observations, we design two generative strategies that guide data selection during training. Finally, we evaluate whether these strategies generalize beyond synthetic conditions by testing on real emotional speech datasets. Through this analysis, we aim to clarify the role of emotion in ASR degradation and demonstrate how targeted data construction can enhance system robustness under emotional variability. Our main findings are:

\begin{enumerate}
\item Synthesized emotional speech increases substitution errors, revealing emotion’s impact on phonetics.
\item Emotionally salient and correctly transcribed samples are more effective for ASR training.
\item Filtered synthetic data improves recognition on real emotional speech without harming neutral performance.
\end{enumerate}

\section{Synthesized Data Analysis and Emotion Impact on ASR System}
\begin{figure}[t]
\centering
\includegraphics[width=8.7cm,height=3.75cm]{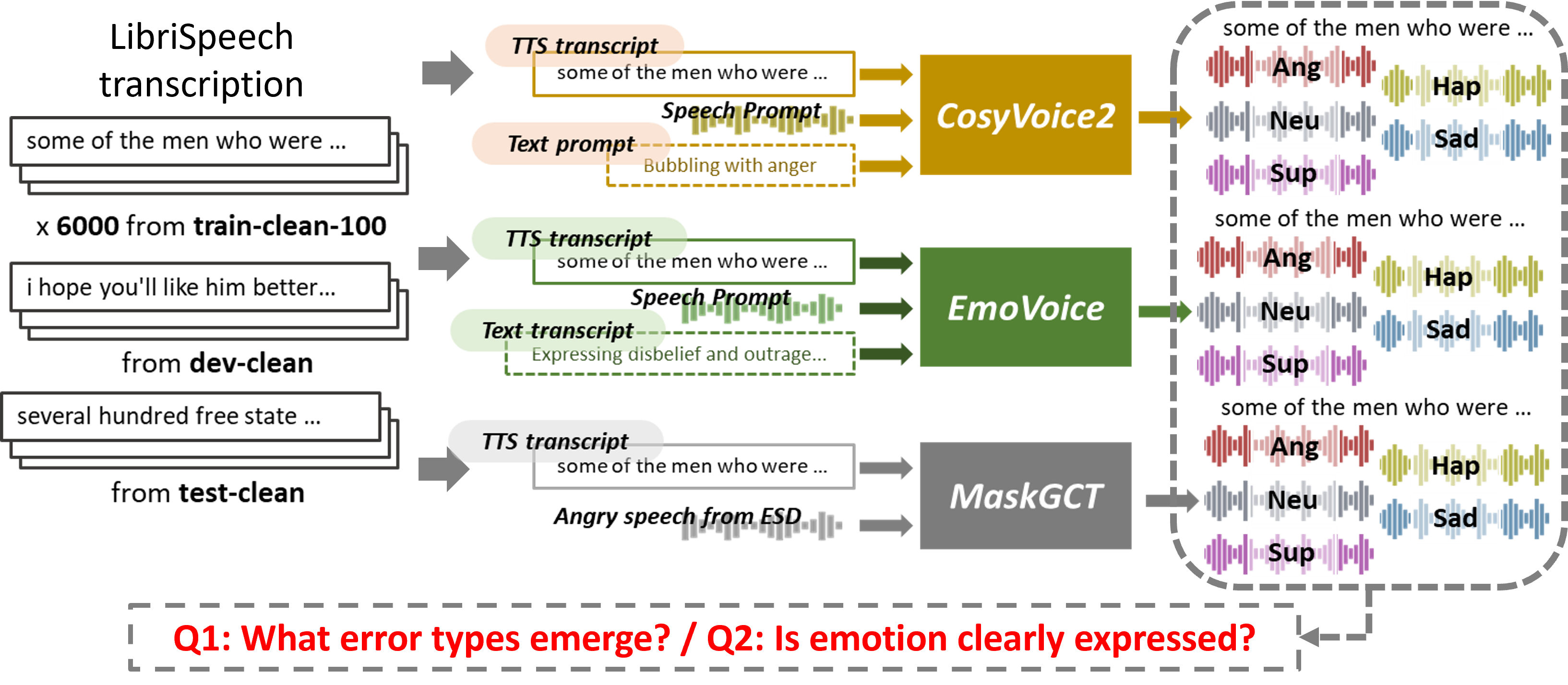}
  \caption{Workflow for synthesized emotional speech analysis.}
  \label{fig:work}
\end{figure}

In this section, we analyze the characteristics of emotional speech synthesized using three state-of-the-art emotion-controllable TTS models. An overview of the analysis workflow is presented in Figure ~\ref{fig:work}. We conduct two main analyses: (1) examining ASR error patterns, including substitution, insertion, and deletion errors, and (2) evaluating the emotional salience of the synthesized speech based on the predicted distributions of dimensional emotion scores.

\subsection{Emotion Controllable TTS Data Synthesis}
To systematically investigate the impact of emotional variability on ASR performance, we control two key aspects in our synthetic speech generation process: transcription content and emotional expression. To minimize lexical variation and isolate the effect of emotion, we restrict the set of transcriptions used for synthesis. Specifically, we sample 6,000 utterances from the LibriSpeech train-clean-100 subset \cite{panayotov2015librispeech}, yielding approximately 20 hours of speech. In addition, the full dev-clean and test-clean subsets are included as additional transcription sources.

Each transcription is synthesized in five representative emotional categories: Angry, Happy, Neutral, Sad, and Surprise. We utilize three state-of-the-art emotion-controllable TTS models to generate the speech data. For each model, the resulting dataset comprises 30,000 utterances for training, 13,470 for development, and 13,055 for testing.



\subsubsection{CosyVoice2 \cite{du2024cosyvoice}}
CosyVoice2 is a multilingual speech synthesis system that performs supervised learning over discrete speech tokens. Its architecture comprises three components: a supervised speech tokenizer, a text-to-speech language model, and a chunk-aware flow-matching module. The system is based on Qwen2.5-0.5B, a causal pre-trained language model with 24 Transformer layers and approximately 0.49 billion parameters. In our experiments, we use the instruction-tuned version of CosyVoice2 without additional fine-tuning. Speech is synthesized by providing a speech prompt, a textual descriptor, and standard linguistic features. To control emotion, we prepend a fixed textual prompt of the form “Bubbling with \{emotion\}” to the input while keeping the original speech prompt unchanged to preserve speaker identity.

\subsubsection{EmoVoice \cite{yang2025emovoice}}
EmoVoice is an emotion-controllable TTS model that enables fine-grained emotional synthesis through natural language prompting. It is also built on Qwen2.5-0.5B but is further trained on an emotion-enriched dataset constructed using GPT-4o and GPT-4o-audio, allowing more nuanced emotional control. In our setup, we use the pretrained EmoVoice 1.5B model to synthesize speech from LibriSpeech transcriptions. To increase emotional diversity, we adopt the “freestyle” prompting strategy proposed by the authors, where a natural language emotion prompt is randomly selected from a predefined set for each utterance. The original speech prompt and transcription are preserved to maintain speaker identity and linguistic consistency. Example prompts are publicly available via the project’s Hugging Face page \footnote{\url{https://huggingface.co/datasets/yhaha/EmoVoice-DB}}.

\subsubsection{MaskGCT \cite{wang2025maskgct}}
MaskGCT is a fully non-autoregressive TTS model that uses masked generative transformers and requires no text-speech alignment or phone-level duration modeling. Unlike LLM-based models, MaskGCT is built on a LLaMA-style Transformer architecture and generates speech conditioned on a semantic speech prompt, which encodes prosodic, stylistic, and emotional information. In our experiments, we use the pretrained MaskGCT model released by the authors. For each synthesized utterance, we randomly select an emotional speech sample from the English subset of the ESD dataset \cite{zhou2021seen} that shares the target emotion (e.g., Angry for Angry synthesis) to serve as the speech prompt, which is then paired with a LibriSpeech transcription to generate emotional speech.


\subsection{Synthesized Data Analysis Methods}
To comprehensively evaluate the quality of synthesized emotional speech, we adopt two complementary analysis strategies: (1) assessing transcription correctness to determine whether the synthesized speech accurately reflects the intended text, and (2) evaluating emotional salience to ensure that the generated speech conveys clear emotional expression rather than emotional neutrality.

\subsubsection{TTS Correctness Analysis}
One standard approach to assess the quality of synthesized speech is to evaluate transcription correctness using Word Error Rate (WER), which captures substitution, deletion, and insertion errors. Analyzing the distribution of these errors in synthesized emotional speech allows us to better understand how emotional content affects ASR transcription fidelity.

We use Qwen2-audio \cite{chu2024qwen2} as the ASR backbone for this analysis. This Speech-LLM architecture integrates Whisper-large-v3 \cite{radford2023robust} as the audio encoder and Qwen-7B \cite{bai2023qwen} as the language model, jointly modeling acoustic and linguistic representations. We select Qwen2-audio due to its state-of-the-art performance on conversational English benchmarks such as Common Voice 15 \cite{ardila2019common}, and its demonstrated ability to capture emotional nuances in speech, which is critical for evaluating ASR performance under affective conditions \cite{feng20_interspeech}.


\subsubsection{Emotion Salience Analysis}
To assess whether the synthesized speech carries sufficient emotional salience, we employ an emotion regression model that estimates Arousal (Act), Valence (Val), and Dominance (Dom) on a 1-to-7 scale, where values near 4 indicate emotional neutrality.

The regressor is implemented as a multitask model based on the WavLM architecture, following the configuration used in the MSP-Podcast Challenge \cite{Goncalves_2024}. It jointly predicts Act, Val, and Dom to capture a multidimensional emotional representation. To prioritize acoustic over lexical features, the model is trained on the BIIC-Podcast dataset \cite{upadhyay2023intelligent}, a 157-hour Chinese podcast corpus with utterance-level annotations in all three emotion dimensions. Although the training domain differs from our synthesized English speech, we directly adopt the original hyperparameters without modification. On the BIIC-Podcast development set, the regressor achieves concordance correlation coefficients (CCC) of 0.68, 0.47, and 0.40 for Act, Val, and Dom, respectively.



\begin{table}[t]
\caption{WER comparison of emotional TTS and LibriSpeech on Qwen2-audio.}
\addtolength{\tabcolsep}{6pt}
\centering
\renewcommand{\arraystretch}{1}
\begin{tabular}{lrrrc}
\hline
WER on Qwen2-Audio     & \multicolumn{1}{l}{Train} & \multicolumn{1}{l}{Dev} & \multicolumn{1}{l}{Test} \\ \hline
LibriSpeech & .0135                     & .0162                   & .0173                    \\
CosyVoice2  & .0316                     & .0350                   & .0357                    \\
EmoVoice    & .1073                     & .0819                   & .0915                    \\
MaskGCT     & .0530                     & .0677                   & .0742                    \\ \hline
\end{tabular}
\label{table:exp1}
\vspace{-2mm}
\end{table}

\begin{table}[t]
\addtolength{\tabcolsep}{0pt}
\centering
\vspace{-2mm}
\caption{Predicted MOS scores (mean $\pm$ standard deviation) for synthesized emotional speech using NISQA.}
\begin{tabular}{lcccc}
\toprule
  & LibriSpeech & CosyVoice2 & EmoVoice & MaskGCT \\
\midrule
MOS & 4.06 $\pm$ 0.55 & 3.79 $\pm$ 0.63 & 3.85 $\pm$ 0.56 & 4.40 $\pm$ 0.42 \\
\bottomrule
\end{tabular}
\label{table:mos}
\vspace{-3mm}
\end{table}

\begin{figure}[t]
\centering
\includegraphics[width=8.7cm,height=3.5cm]{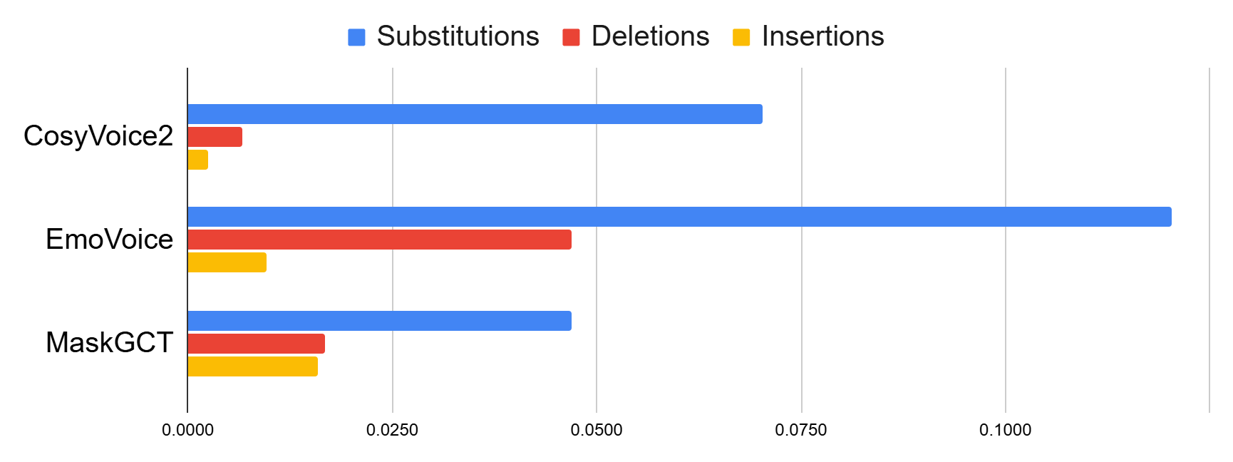}
\vspace{-2mm}
  \caption{Distribution of word error types in the training set of emotional TTS speech. }
  \label{fig:sdi}
  \vspace{-7mm}
\end{figure}

\begin{figure}[t]
\centering
\includegraphics[width=8.7cm,height=3cm]{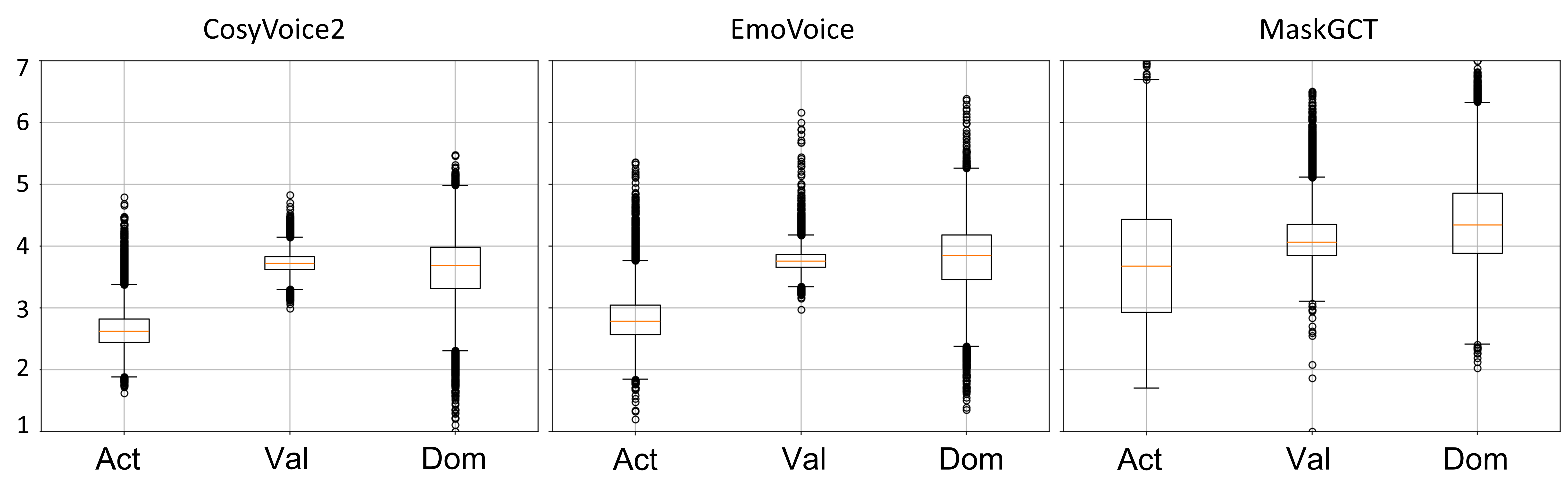}
  \vspace{-3mm}
  \caption{Box plots of emotion distributions in the training set of emotional TTS speech.}
  \vspace{-4mm}
  \label{fig:emo_dist}
\end{figure}

\subsection{Result and Discussion}
Table~\ref{table:exp1} shows that all emotional TTS datasets lead to significantly higher WERs compared to the original LibriSpeech transcriptions, which have an average WER of 1.57\% across splits. This supports prior findings that emotional utterances are more error-prone in ASR than neutral ones \cite{catania2019automatic}. Among the synthesized datasets, CosyVoice2 yields the lowest WER at 3.74\%, while MaskGCT and EmoVoice produce higher error rates at 6.50\% and 9.37\% in average, respectively. These results confirm that synthesized emotional speech presents additional challenges for ASR. The relatively low WER of CosyVoice2 suggests greater consistency and transcriptional clarity, whereas the high WER of EmoVoice likely stems from more expressive and variable prosody that increases recognition difficulty.

To ensure that transcription errors are not caused by degraded speech quality, we evaluate the synthesized utterances in training set using the NISQA model, a non-intrusive speech quality assessment metric \cite{mittag21_interspeech}. As shown in Table~\ref{table:mos}, all three emotional TTS systems generate high-quality audio with the average of Mean Opinion Score (MOS) estimates above 3.7, with MaskGCT achieving particularly strong results (4.40). These findings suggest that the elevated WERs are unlikely due to poor signal quality, and instead reflect challenges introduced by emotional expressiveness in the generated speech.

Figure~\ref{fig:sdi} illustrates the distribution of substitution, insertion, and deletion errors on three synthesized training sets. Error rates are first normalized by the word length of each sentence and then averaged across the training set. Substitution errors are clearly dominant across all TTS systems, while insertions and deletions remain less frequent. This indicates that affective speech tends to cause semantic-level misrecognitions rather than structural disruptions. Notably, both EmoVoice and MaskGCT exhibit higher insertion and deletion rates than CosyVoice2, suggesting a higher tendency toward hallucination-related phenomena \cite{10890943}, such as skipped or repeated words.

Figure~\ref{fig:emo_dist} presents the box plot distributions of Arousal (Act), Valence (Val), and Dominance (Dom) scores predicted by the emotion regressor. CosyVoice2 and EmoVoice show skewed Arousal distributions concentrated below 4, indicating weak emotional activation. MaskGCT, by contrast, produces a broader spread across the Act range. Valence distributions across all models are narrow with limited variability, although MaskGCT again demonstrates comparatively higher diversity. These results suggest that only a portion of synthesized utterances exhibit clear emotional salience.

In summary, our analysis reveals two key issues with synthesized emotional data: elevated ASR error rates, primarily due to substitution errors, and limited emotional expressiveness in a substantial portion of utterances. These findings highlight the importance of selectively leveraging high-quality, emotionally salient samples. In the next section, we present two generative strategies, informed by transcription correctness and emotion salience, to construct more effective synthetic training sets and assess their impact on ASR performance.

\section{Generative Strategies and Comparison on Synthesized Data}
Motivated by the findings in the previous section, we develop two generative strategies to improve the effectiveness of synthesized emotional speech for ASR training. The first strategy selects utterances based on transcription correctness, and the second identifies samples with strong emotional salience. These strategies are designed to address the specific issues observed in our earlier analysis, including high substitution error rates and limited emotional expressiveness, by constructing targeted training subsets that emphasize both linguistic accuracy and emotional clarity.


\subsection{Generative Strategy}
\subsubsection{TTS Correctness (TTS-G)}
To address emotion-induced recognition errors, we define a generative strategy that selects synthesized utterances which introduce substitution errors, identified in our earlier analysis as the dominant error type. For each synthesized utterance derived from a LibriSpeech transcription, we compare its ASR error profile to that of its original counterpart using the same text. An utterance is retained only if it results in a higher number of substitutions while maintaining equal or fewer deletion and insertion errors. Formally, the selected subset $\mathcal{U}_T$ is defined as:

\begin{equation}
\mathcal{U}_T = \left\{ u \in \mathcal{U} \,\middle|\, 
\begin{aligned}
&S_s(u) > S_o(u) \;\land \\
&D_s(u) \leq D_o(u) \;\land \\
&I_s(u) \leq I_o(u)
\end{aligned}
\right\}
\end{equation}

where $S(u)$, $D(u)$, and $I(u)$ denote the number of substitutions, deletions, and insertions respectively, and the subscripts $s$ and $o$ refer to synthesized and original utterances.

\subsubsection{Emotion Salience (EMO-G)}
To ensure that the speech in the augmented data has sufficient emotional expression, we define a generative strategy based on emotion salience. For each synthesized utterance, we compute the predicted dimensional emotion scores and retain it if at least one of the three dimensions—Arousal (Act), Valence (Val), or Dominance (Dom)—deviates by more than one standard deviation from its corresponding mean. Such deviations indicate a departure from emotional neutrality and reflect stronger emotional expression. The selected subset $\mathcal{U}_E$ is defined as:

\begin{equation}
\mathcal{U}_E = \left\{ u \in \mathcal{U} \,\middle|\, 
\begin{aligned}
&|\text{Act}_u - \mu_{\text{Act}}| > \sigma_{\text{Act}} \;\lor \\
&|\text{Val}_u - \mu_{\text{Val}}| > \sigma_{\text{Val}} \;\lor \\
&|\text{Dom}_u - \mu_{\text{Dom}}| > \sigma_{\text{Dom}}
\end{aligned}
\right\}
\end{equation}

where $\mu$ and $\sigma$ denote the mean and standard deviation across the synthesized dataset.

\subsubsection{Combined Generative Strategy (TTS-EMO-G)}
To further enhance data quality, we apply both the TTS correctness and emotion salience criteria in combination. An utterance is retained only if it satisfies both conditions. This joint strategy ensures that selected samples both exhibit strong emotional expression and introduce meaningful recognition errors, increasing their potential value for ASR training. We refer to this combined approach as TTS-EMO-G in our experiments.

\subsection{ASR Model Fine-tuning Details}
To evaluate the effectiveness of our generative strategies, we fine-tune the pretrained Qwen2-audio-7B model using the training portions of the TTS-EMO-G subsets, which are constructed following the original LibriSpeech train/dev/test splits for each TTS system (CosyVoice2, EmoVoice, and MaskGCT). Fine-tuning is performed independently for each system using only its synthesized training set, with early stopping based on the lowest development loss from the corresponding dev set. ASR transcriptions are generated using the official Qwen2-audio inference prompt: “Detect the language and recognize the speech:”

Training is conducted using ms-swift v2.6.1~\cite{zhao2025swift}, a flexible framework for tuning large language model-based systems. We adopt a supervised fine-tuning approach in which only a small portion of the audio encoder (AudioEnc) is updated, while the rest of the model remains frozen. Specifically, we allow gradient updates on the final 0.1\% of encoder parameters, amounting to approximately 1.27 million trainable parameters.

We use the AdamW optimizer from PyTorch 2.4.1~\cite{paszke2019pytorch}, with a learning rate of $10^{-5}$. All hyperparameters follow the default ms-swift configuration, including a warm-up ratio of 0.05, gradient accumulation steps of 16, weight decay of 0.1, and a maximum gradient norm of 1. The model is trained for 5 epochs using cross-entropy loss. Each fine-tuning run takes approximately 72 to 96 hours on two NVIDIA A100 or V100 GPUs. Further implementation details are available in the official ms-swift2 documentation\footnote{\url{https://swift.readthedocs.io/en/stable/}}.

\begin{table}[t] 
\caption{WER across subsets selected by different generative strategies.}
\addtolength{\tabcolsep}{10pt}
\centering
\renewcommand{\arraystretch}{0.75}
\begin{tabular}{@{}cccc@{}}
\toprule
WER on Qwen2-Auido             & Train & Dev   & Test  \\ \midrule
\multicolumn{4}{c}{CosyVoice2}          \\ \midrule
Vanilla         & .0316 & .0350 & .0357 \\
EMO-G         & .0332 & .0356 & .0364 \\
TTS-G         & .0333 & .0367 & .0369 \\
TTS-EMO-G     & .0358 & .0406 & .0401 \\ \midrule
\multicolumn{4}{c}{EmoVoice}            \\ \midrule
Vanilla         & .1073 & .0819 & .0915 \\
EMO-G         & .1059 & .0815 & .1024 \\
TTS-G         & .0434 & .0447 & .0468 \\
TTS-EMO-G     & .0460 & .0466 & .0504 \\ \midrule
\multicolumn{4}{c}{MaskGCT}             \\ \midrule
Vanilla         & .0530 & .0677 & .0742 \\
EMO-G         & .0494 & .0622 & .0901 \\
TTS-G         & .0381 & .0464 & .0447 \\
TTS-EMO-G     & .0396 & .0508 & .0467 \\ \bottomrule
\end{tabular}
\label{table:exp2}
\end{table}

\begin{table}[t]
\caption{WER on synthesized test sets using TTS-EMO-G with AudioEnc fine-tuning.}
\addtolength{\tabcolsep}{-2pt}
\centering
\begin{tabular}{@{}ccccc@{}}
\toprule
\multicolumn{1}{l}{} & Qwen2-Audio & CosyVoice2      & EmoVoice & MaskGCT         \\ \midrule
CosyVoice2-Test      & .0401      & .0196          & .0291   & .0285          \\
EmoVoice-Test        & .0504      & .0351          & .0282   & .0347          \\
MaskGCT-Test         & .0467      & .0414          & .0991   & .0284          \\
LibriSpeech       & .0173      & .0169          & .0207   & .0184          \\ \midrule
Avg                  & .0386      & .0283 & .0443   & \textbf{.0275} \\ \bottomrule
\end{tabular}
\label{table:exp3}
\vspace{-4mm}
\end{table}

\subsection{Results and Discussion}
To determine whether the proposed generative strategies provide meaningful benefits, we first evaluate their direct impact on the pretrained Qwen2-audio model without any fine-tuning. Table~\ref{table:exp2} presents the WER results on subsets selected by three strategies: (1) TTS-G, which filters utterances based on transcription correctness; (2) EMO-G, which selects utterances with strong emotional salience based on dimensional emotion scores; and (3) TTS-EMO-G, which combines both criteria to retain samples that are both transcriptionally informative and emotionally expressive. Compared to the original synthesized dataset (Vanilla), these strategies yield notable variations across the training, development, and testing splits. Notably, TTS-EMO-G consistently results in higher WERs than TTS-G, with average increases of 0.32\%, 0.27\%, and 0.26\% for CosyVoice2, EmoVoice, and MaskGCT, respectively. This increase is expected, as TTS-EMO-G emphasizes samples with greater acoustic variability and emotional expressiveness, introducing additional complexity for ASR models.

Table~\ref{table:exp3} further reports the results after fine-tuning under the TTS-EMO-G setting. Evaluations are conducted on the synthesized test sets from each emotional TTS model, as well as on the original LibriSpeech test-clean set. With AudioEnc fine-tuning, most configurations yield consistent WER reductions relative to the pretrained baseline, except for EmoVoice when tested on the MaskGCT set. Crucially, performance on LibriSpeech remains largely stable across all conditions, confirming that incorporating emotionally expressive synthetic data does not degrade recognition accuracy on clean, neutral speech. Specifically, WER improves by 0.34\% for EmoVoice and 0.11\% for MaskGCT, with only a slight increase of 0.04\% observed for CosyVoice2. These results demonstrate that the TTS-EMO-G strategy can effectively mitigate substitution errors introduced by emotional variability while preserving general ASR performance.

Among the three emotional TTS models, MaskGCT achieves the best overall performance, with an average WER of 2.75\%, while EmoVoice shows the least improvement, averaging 4.43\%. MaskGCT’s advantage is attributed to its more balanced emotional distribution, as illustrated in Figure~\ref{fig:emo_dist}. Compared to CosyVoice2 and EmoVoice, MaskGCT exhibits fewer outliers and more centralized emotion scores across Arousal, Valence, and Dominance, closer to the neutral midpoint of 4. This balance likely contributes to more stable acoustic patterns and greater ASR robustness under affective conditions. These findings highlight the effectiveness of the TTS-EMO-G strategy in controlled synthetic environments. In the following section, we evaluate whether similar improvements can be achieved on real-world emotional speech.

\section{Real Emotion Data Evaluation}
To examine whether our generative strategies generalize to real emotional speech, we evaluate the ASR models fine-tuned on the synthesized subsets introduced in the previous section. Specifically, we assess models trained using the three generative strategies: TTS-G, EMO-G, and TTS-EMO-G, on three benchmark datasets: MSP Podcast Test1, Test2, and IEMOCAP. These datasets are used only for evaluation, and no real emotional speech is seen during training. This setup allows us to determine whether models trained solely on synthetic data can generalize to real world affective conditions and to identify which emotional regions benefit from our approach.

\subsection{Real Speech Emotion Datasets}
\subsubsection{MSP-Podcast}
We use the MSP-Podcast corpus v1.11 \cite{lotfian2017building} as the real emotional speech benchmark to evaluate our fine-tuned ASR models. This dataset contains emotionally rich speech collected from online platforms, spanning diverse speakers and topics. The recordings are segmented into utterances (16 kHz, mono) and annotated with transcriptions, speaker identities, and dimensional emotion ratings for Arousal, Valence, and Dominance. Each dimension is scored on a scale from 1 to 7. Following the official evaluation protocol, we use the predefined Test 1 and Test 2 splits, which contain 30,647 and 14,815 utterances, respectively, totaling 70.94 hours of speech.


\subsubsection{IEMOCAP}
We also evaluate model performance on the IEMOCAP corpus \cite{busso2008iemocap}, a widely used emotional speech dataset consisting of dyadic conversations recorded in a controlled lab setting. The corpus includes five sessions with ten speakers, and each utterance is annotated with a transcript and dimensional emotion scores in Arousal, Valence, and Dominance. Each dimension is rated on a scale from 1 to 5. In this study, we use the entire corpus for evaluation, comprising 10,044 utterances and approximately 12 hours of speech.

\begin{table*}[t]
\caption{WER on MSP Test1, Test2, and IEMOCAP using different generative strategies with AudioEnc fine-tuning.}
\addtolength{\tabcolsep}{-2pt}
\centering
\begin{tabular}{@{}ccccccccccccc@{}}
\toprule
\multicolumn{1}{l|}{}            & \multicolumn{4}{c|}{MSP Test1}                                & \multicolumn{4}{c|}{MSP Test2}                                & \multicolumn{4}{c}{IEMOCAP}              \\ \midrule
\multicolumn{1}{c|}{Qwen2-Audio} & \multicolumn{4}{c|}{.1578}                                    & \multicolumn{4}{c|}{.1541}                                    & \multicolumn{4}{c}{.0927}                \\ \midrule
\multicolumn{1}{l|}{}            & Vanilla & EMO-G & TTS-G & \multicolumn{1}{c|}{TTS-EMO-G}      & Vanilla & EMO-G & TTS-G & \multicolumn{1}{c|}{TTS-EMO-G}      & Vanilla & EMO-G & TTS-G & TTS-EMO-G      \\ \midrule
\multicolumn{1}{c|}{CosyVoice2}  & .1685   & .1725 & .1692 & \multicolumn{1}{c|}{.1601}          & .1450   & .1449 & .1458 & \multicolumn{1}{c|}{\textbf{.1442}} & .1044   & .0891 & .0956 & .0956          \\
\multicolumn{1}{c|}{EmoVoice}    & .2271   & .2092 & .1611 & \multicolumn{1}{c|}{.1970}          & .2297   & .1818 & .1453 & \multicolumn{1}{c|}{.1850}          & .1380   & .1399 & .1090 & .1336          \\
\multicolumn{1}{c|}{MaskGCT}     & .1620   & .1743 & .1577 & \multicolumn{1}{c|}{\textbf{.1560}} & .1528   & .1519 & .1598 & \multicolumn{1}{c|}{.1517}          & .1116   & .0990 & .0892 & \textbf{.0870} \\ \bottomrule
\end{tabular}
\label{table:exp4}
\end{table*}

\subsection{Result and Discussion}
Table~\ref{table:exp4} reports the performance of all generative strategies evaluated on MSP-Podcast Test1, Test2, and IEMOCAP. The highest WER improvements observed are 0.18\%, 0.99\%, and 0.57\% on each dataset, respectively. Among the three strategies, TTS-EMO-G consistently achieves the best performance across all benchmarks. Notably, our model was not trained on any of these datasets and had no exposure to their emotional distributions. Despite this, it demonstrates measurable improvements, indicating strong generalization. Furthermore, the training transcriptions were originally well-recognized by the baseline ASR system (WER of 1.35\% as shown in Table~\ref{table:exp1}), suggesting that the observed gains stem from emotion-induced variability rather than lexical differences. These findings confirm that the proposed generative strategies are effective not only on synthetic emotional data but also generalize well to real-world emotional speech.

Among the three TTS methods, MaskGCT achieves the best performance on IEMOCAP, consistent with earlier findings that its well-balanced emotional distribution improves ASR robustness. On MSP Test1 and Test2, the top-performing systems are based on MaskGCT and CosyVoice2, respectively. MSP Test1, curated using emotion prediction outputs, appears to benefit from MaskGCT's expressive range. In contrast, MSP Test2 contains more emotionally neutral content, where CosyVoice2's stable synthesis and lower hallucination rate (as shown in Table~\ref{table:exp1}) lead to better performance. As an additional observation, the combination of MaskGCT and TTS-EMO-G consistently improved over the pretrained baseline across all test sets, suggesting its robustness to varying emotional distributions.

To further understand where performance gains occur, we analyze WER changes across the emotional space using dimensional annotations from MSP Test2 and IEMOCAP. For MSP Test2, utterances are grouped by their rounded Arousal, Valence, and Dominance scores on a 1--7 scale, and WER is computed for each group across three 2D planes: (Act, Val), (Val, Dom), and (Act, Dom). For IEMOCAP, we adopt a coarse-grained 3$\times$3 binning strategy. Each emotional dimension, originally scored from 1 to 5, is categorized into three levels: low ($\leq$2.5), medium ($>$2.5 and $\leq$3.5), and high ($>$3.5). Utterances are grouped accordingly, and WER is calculated across the same three 2D planes to visualize ASR performance trends under varying emotional intensities.

\begin{figure}[t]
\centering
\includegraphics[width=8cm,height=5.4cm]{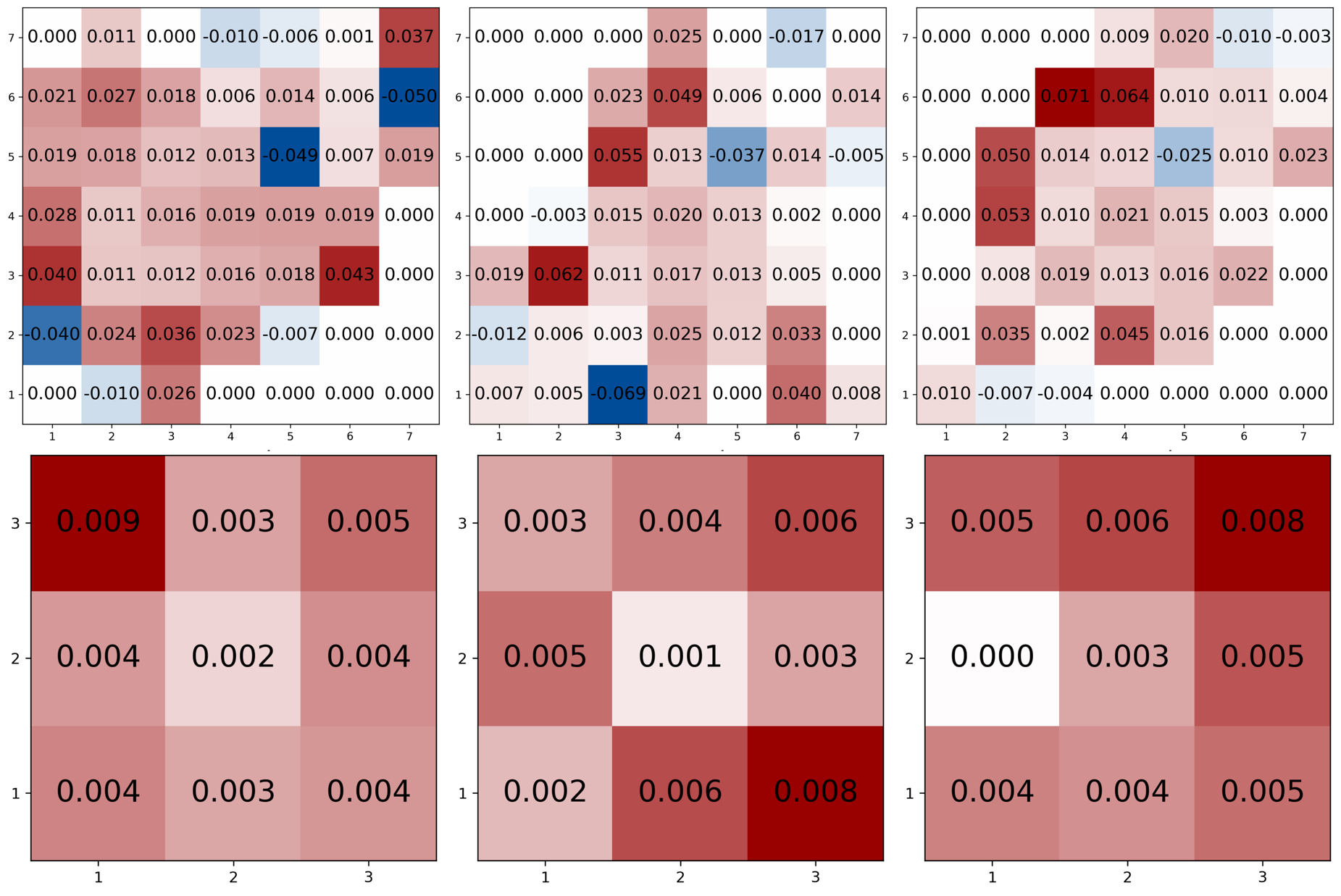}
  \caption{Left to right: (Act, Val), (Val, Dom), and (Act, Dom) planes showing WER difference patterns for MSP-Test 2 (top) and IEMOCAP (bottom).}
  \label{fig:emo_plane}
  \vspace{-4mm}
\end{figure}

Figure~\ref{fig:emo_plane} visualizes the WER differences between the pretrained model and the best fine-tuned system. Red indicates improved WER, blue indicates degradation. Improvements are most pronounced at the emotional extremes, confirming that our strategies are especially effective for enhancing ASR performance under emotionally expressive conditions.

\section{Discussion and Conclusion}
This study investigates the impact of synthesized emotional speech on ASR performance. Analysis of three emotional TTS models shows that emotional variability mainly causes substitution errors, suggesting that affective modulation affects phonetic realization more than lexical structure. We also observe significant variation in emotional expressiveness and generative stability across TTS systems, influencing both recognition accuracy and emotion distribution.

To address these challenges, we propose two generative strategies: one based on transcription correctness and the other on emotional salience—that guide the selection of emotionally rich yet linguistically reliable training subsets. Fine-tuning with the combined strategy (TTS-EMO-G) yields consistent WER improvements on both synthetic and real emotional datasets, particularly in highly expressive regions, while maintaining performance on neutral speech. These results demonstrate the feasibility of leveraging high-quality synthetic data for emotion-aware ASR training.

This approach relies on well-trained emotional TTS systems and accurate dimensional emotion regression, limiting applicability in low-resource or cross-lingual settings. Since all synthesized data are derived from LibriSpeech transcripts, domain mismatches between synthetic and spontaneous speech may hinder generalization. The analysis also focuses on dimensional emotion representation (arousal, valence, dominance) rather than discrete categories; given that some emotions, such as sadness, may be acoustically closer to neutral, incorporating per-emotion analysis could offer complementary insights.

Beyond acoustic adaptation, incorporating lexical cues and contextual semantics may enhance robustness under emotionally complex conditions and broaden applications in multilingual and low-resource scenarios. Future work could explore language-agnostic augmentation techniques, finer-grained emotion control, and adaptive data selection strategies informed by speaker or context. Investigating lexical items particularly susceptible to recognition errors under different emotional conditions may enable targeted augmentation or model adaptation for these high-risk words.

\bibliographystyle{IEEEtran}
\bibliography{refs}

\end{document}